\documentclass[sigconf]{acmart}
\AtBeginDocument{%
  \providecommand\BibTeX{{%
    \normalfont B\kern-0.5em{\scshape i\kern-0.25em b}\kern-0.8em\TeX}}}

\copyrightyear{2024} 
\acmYear{2024} 
\setcopyright{rightsretained} 
\acmConference[ICSE '24]{2024 IEEE/ACM 46th International Conference on Software Engineering}{April 14--20, 2024}{Lisbon, Portugal}
\acmBooktitle{2024 IEEE/ACM 46th International Conference on Software Engineering (ICSE '24), April 14--20, 2024, Lisbon, Portugal}
\acmDOI{10.1145/3597503.3623298}
\acmISBN{979-8-4007-0217-4/24/04}

\usepackage{soul}
\usepackage{multirow}
\usepackage{epstopdf}
\usepackage{hyperref}
\usepackage{listings}
\usepackage{fancybox}
\usepackage{graphicx}
\usepackage{subcaption}
\usepackage{color}
\usepackage{paralist}
\usepackage{ragged2e}
\usepackage{enumitem}
\usepackage{algorithm}
\usepackage{algorithmic}
\usepackage{booktabs}
\usepackage{pgf-pie}
\usepackage[utf8]{inputenc}
\usepackage{listings}

\usepackage[framemethod=TikZ]{mdframed}
\usepackage{tcolorbox}
\makeatletter
\newcommand{\mybox}[1]{%
  \setbox0=\hbox{#1}%
  \setlength{\@tempdima}{\dimexpr\wd0+13pt}%
  \begin{tcolorbox}[boxrule=0.5pt, colback=white, arc=4pt,
      left=4pt,right=4pt,top=4pt,bottom=4pt,boxsep=0pt]
    #1
  \end{tcolorbox}
}
\usepackage{tikz}
\usepackage{pgfplots}
\usetikzlibrary{pgfplots.statistics,calc}
\usepackage{color}
\usepackage{xcolor}
\usepackage{array}
\usepackage{amsmath}

\usepackage{amssymb}
\usepackage{centernot}
\usepackage{xspace}
\usepackage{url}
\usepackage{bbding}
\usepackage{verbatim}
\usepackage{wrapfig}
\usepackage{tabularx}
\usepackage{colortbl}
\definecolor{javared}{rgb}{0.6,0,0} 
\definecolor{javagreen}{rgb}{0.25,0.5,0.35} 
\definecolor{javapurple}{rgb}{0.5,0,0.35} 
\definecolor{javadocblue}{rgb}{0.25,0.35,0.75} 

\newcommand{\tool}{CrashTranslator}

\usepackage[normalem]{ulem} 
\newcommand\hly{\bgroup\markoverwith
  {\textcolor{yellow!10}{\rule[-.5ex]{2pt}{2.5ex}}}\ULon}
\newcommand\hlc{\bgroup\markoverwith
  {\textcolor{cyan!10}{\rule[-.5ex]{2pt}{2.5ex}}}\ULon}
\newcommand\hlp{\bgroup\markoverwith
  {\textcolor{purple!10}{\rule[-.5ex]{2pt}{2.5ex}}}\ULon}
\newcommand\hlo{\bgroup\markoverwith
  {\textcolor{orange!10}{\rule[-.5ex]{2pt}{2.5ex}}}\ULon}
\newcommand\hlg{\bgroup\markoverwith
  {\textcolor{green!10}{\rule[-.5ex]{2pt}{2.5ex}}}\ULon}
\newcommand\hlgg{\bgroup\markoverwith
  {\textcolor{brown!10}{\rule[-.5ex]{2pt}{2.5ex}}}\ULon}
  
\begin{document}

\title{CrashTranslator: Automatically Reproducing Mobile Application Crashes Directly from Stack Trace}


\author{Yuchao Huang}
\authornote{Also With State Key Laboratory of Intelligent Game, Institute of Software, CAS}
\authornote{Also With Science and Technology on Integrated Information System Laboratory, Institute of Software, CAS}
\authornote{Also With University of Chinese Academy of Sciences}
\email{yuchao2019@iscas.ac.cn}
\affiliation{%
  \institution{Institute of Software, Chinese Academy of Sciences}
  \city{Beijing}
  \country{China}
}

\author{Junjie Wang}
\authornotemark[1]
\authornotemark[2]
\authornotemark[3]
\authornote{Corresponding author}
\email{junjie@iscas.ac.cn}
\affiliation{%
  \institution{Institute of Software, Chinese Academy of Sciences}
  \city{Beijing}
  \country{China}
}

\author{Zhe Liu}
\authornotemark[1]
\authornotemark[2]
\authornotemark[3]
\email{liuzhe181@mails.ucas.edu.cn}
\affiliation{%
  \institution{Institute of Software, Chinese Academy of Sciences}
  \city{Beijing}
  \country{China}
}

\author{Yawen Wang}
\authornotemark[1]
\authornotemark[2]
\authornotemark[3]
\email{yawen2018@iscas.ac.cn}
\affiliation{%
  \institution{Institute of Software, Chinese Academy of Sciences}
  \city{Beijing}
  \country{China}
}

\author{Song Wang}
\email{wangsong@yorku.ca}
\affiliation{%
  \institution{York University}
  \city{Toronto}
  \country{Canada}
}

\author{Chunyang Chen}
\email{Chunyang.chen@monash.edu}
\affiliation{%
  \institution{Monash University}
  \city{Melbourne}
  \country{Australia}
}

\author{Yuanzhe Hu}
\authornotemark[1]
\authornotemark[2]
\authornotemark[3]
\email{yuanzhe@iscas.ac.cn}
\affiliation{%
  \institution{Institute of Software, Chinese Academy of Sciences}
  \city{Beijing}
  \country{China}
}

\author{Qing Wang}
\authornotemark[1]
\authornotemark[2]
\authornotemark[3]
\authornotemark[4]
\email{wq@iscas.ac.cn}
\affiliation{%
  \institution{Institute of Software, Chinese Academy of Sciences}
  \city{Beijing}
  \country{China}
}

\begin{abstract}

Crash reports are vital for software maintenance since they allow the developers to be informed of the problems encountered in the 
mobile application.
Before fixing, developers need to reproduce the crash, which is an extremely time-consuming and tedious task. Existing studies conducted the automatic crash reproduction with the natural language described reproducing steps.
Yet we find a non-neglectable portion of crash reports only contain the stack trace when the crash occurs. Such stack-trace-only crashes merely reveal the last GUI page when the crash occurs, and lack step-by-step guidance. Developers tend to spend more effort in understanding the problem and reproducing the crash, and existing techniques cannot work on this, thus calling for a greater need for automatic support. This paper proposes an approach named {\tool} to automatically reproduce mobile application crashes directly from the stack trace. It accomplishes this by leveraging a pre-trained Large Language Model to predict the exploration steps for triggering the crash, and designing a reinforcement learning based technique to mitigate the inaccurate prediction and guide the search holistically. We evaluate {\tool} on 75 crash reports involving 58 popular Android apps, and it successfully reproduces 61.3\% of the crashes, outperforming the state-of-the-art baselines by 109\% to 206\%. Besides, the average reproducing time is 68.7 seconds, outperforming the baselines by 302\% to 1611\%. We also evaluate the usefulness of {\tool} with promising results.
\end{abstract}

\ccsdesc[500]{Software and its engineering~Software maintenance tools}

\keywords{Bug reproduction, Stack trace, Mobile application testing}


\maketitle

\section{INTRODUCTION}
\label{sec_introduction}

New mobile applications are being developed and released continuously via app stores since the market for mobile devices is both growing and diversifying. 
Recent statistics show that more than 5 million mobile apps are available in popular marketplaces like Apple App Store and Google Play Store, for over 140 billion downloads in 2022 and 12 million mobile developers are maintaining them \cite{app_statistics}. 
As developers add more features and capabilities to their apps to make them more competitive, the corresponding increase in app complexity has made testing and
maintenance activities more challenging. 
The competitive app marketplace has also made these activities quite important for an app's success. 
As shown in a survey, 88\% of app users would abandon an app if they repeatedly encountered a functionality issue \cite{applause}. 
This motivates developers to identify and resolve issues rapidly, or risk losing users otherwise.

An important mechanism for ensuring app quality is the online bug reporting systems, e.g., GitHub Issue Tracker \cite{github_issue_tracker}, Bugzilla \cite{bugzilla}, Google Code Issue Tracker \cite{google_issue_tracker}. 
These systems enable users to create bug reports in which they can describe their observed failure; developers can then use this information to help debug their apps. 
These bug reports are becoming a non-neglectable source of information for improving app quality and user satisfaction.

Once developers receive a bug report, one of the first steps to debugging the reported issue is to reproduce the issue following the reproducing steps. 
There are several existing studies focusing on crash\footnote{Existing studies focus on the crash reports since they have the observable oracle compared with other bugs, and following them, this work also focuses on the crash reproduction, and we will use crash and bug interchangeably.} reproduction from the natural language described reproducing steps \cite{zhao2019ReCDroid, zhao2022recdroid+, zhang2023automatically}.
They typically apply natural language processing techniques to match the reproducing steps with the app's GUI events (i.e., operations on GUI widgets, e.g., clicking the \textit{search} button of an app), and employ guided exploration strategies with the matched information for bug reproduction. 
However, not all crash submitters would strictly follow the report template to provide the reproducing steps when reporting the crash. 

Our motivational study (Section \ref{sec_movitation}) reveals that a non-neglectable portion (20.2\%) of crash reports contain only the stack trace.
Such stack-trace-only reports can be submitted by crash reporting tools such as Crashlytics \cite{crashlytics}, which automatically collects crash logs and upload them, and this is also the commonly-used practice in large commercial software.
These reports can also be submitted by the app users who accidentally trigger the crash yet fail to figure out the reproducing steps. 
Due to the insufficient information provided in these reports, developers tend to spend extra effort in understanding and reproducing the issues, which brings in a longer fixing duration, i.e., the average fixing duration of these stack-trace-only reports is 26\% larger than the crash reports with reproducing steps. 
Besides, the aforementioned existing approaches would fail to work on stack-trace-only reports.
This further implies the necessity for the automatic crash reproduction approach directly from stack traces. 

Existing studies on stack trace analysis relate with stack trace similarity \cite{khvorov2021s3m, vasiliev2020tracesim, rodrigues2022tracesim}, fault localization from stack trace \cite{wu2014crashlocator, wong2014boosting, moreno2014use, gong2014locating, gu2019does},  test code generation from stack trace \cite{rossler2013reconstructing, chen2014star, xuan2015crash, soltani2016evolutionary, soltani2017guided, soltani2020benchmark}, duplicate crash reports detection with stack trace \cite{dang2012rebucket, rodrigues2022fast}, etc. 
Although these approaches can facilitate the understanding and analysis of the stack trace, none of them can tackle the problem of automatic crash reproduction from the stack trace. 
There are two challenges in the automatic reproduction of these crashes. 

First, as mentioned above, existing studies generally utilize the textual-described reproducing steps for crash reproduction, yet stack-trace-only crashes lack step-by-step guidance. 
In other words, the stack trace does not record the exploration sequence from the entry page to the crash-occurring page, while the most useful information might be the last GUI page when the crash occurs. 
However, there can be 1 to 8 exploration steps for reaching the last crash-occurring GUI page (based on our experimental data), which can be quite inefficient if explored randomly. 
Although the static or dynamic analysis techniques \cite{gator, storydistiller, iccbot} can infer the transition between activities and plan the exploration path, yet they can be quite incomplete or inaccurate \cite{yan2022comprehensive}. 
Second, even when the last crash-occurring page is reached, it may still need certain interactions with the app to finally trigger the crash, e.g., clicking a certain button. 
Nevertheless, there can be an average of 6.6 interactive widgets on a GUI page (based on our experimental data) and the stack trace does not implicitly provide which widget to interact with for crash triggering, which further complicates the automated crash reproduction problem. 

Nevertheless, we also find two clues for facilitating the reproduction of stack-trace-only crashes. 
The first is the last crash-occurring GUI page before triggering the crash, which offers the target for the planned exploration. 
The second is the involved APIs in the programming code when conducting the operations in the crash-occurring page, which can help find the widget with which to interact so the crash can finally be triggered.

Motivated by these clues, we propose an approach named {\tool} to automatically reproduce mobile application crashes directly from the stack trace. It accomplishes this by leveraging a pre-trained Large Language Model (LLM) to predict the exploration steps for triggering the crash, and designing a reinforcement learning based technique to mitigate the inaccurate prediction and guide the search holistically.

In detail, we first extract crash-related information from the stack trace, i.e., the crash-occurring GUI page and crash-involved APIs. Second, we design three scorers to assign the exploration priority for each GUI widget on the current page: 
1) Page reaching scorer, which leverages the LLM to choose the GUI widget that may lead to the crash-occurring page; 2) Widget hitting scorer, which utilizes the heuristic method to find the crash-triggering widget by matching the crash-involved APIs; 3) Exploration optimization scorer, which assigns scores based on previous interaction records of reinforcement learning technique, 
in order to bridge the gap by inaccurate prediction of the first two scorers and plan the exploration holistically. {\tool} selects the GUI widget based on the three scorers and continues the process iteratively until the target crash is triggered. It finally generates the replay script (for direct replay) and the textual-described reproducing steps with the step-by-step image instructions (for facilitating understanding).

To evaluate the effectiveness and efficiency of our approach, we run {\tool} on 75 crash reports collected from 58 popular Android apps involving three datasets. {\tool} successfully reproduces 61.3\% (46/75) of the crashes, which outperforms the state-of-the-art baselines by 109\% to 206\%. 
Besides, the average reproducing time is 68.7 seconds, outperforming the baselines by 302\% to 1611\%.
Furthermore, the results also show that both the designed page reaching scorer and widget hitting scorer greatly contribute to the reproduction performance. 
The usefulness evaluation shows that {\tool}'s generated reproducing steps can make the crashes easily reproduced (215\% faster).

The contributions of this paper are as follows:
\begin{itemize}
    \item \textbf{Dimension.} The first work of the automatic crash reproduction of mobile applications directly from the stack trace.

    \item \textbf{Technique.} An automatic approach {\tool} for stack-trace-only crash reproduction by leveraging the LLM to predict the exploration steps for triggering the crash, and designing a reinforcement learning based technique to mitigate the inaccurate prediction and guide the search holistically.

    \item \textbf{Evaluation.} Experimental evaluation of effectiveness, efficiency and usefulness of {\tool} with promising performance, outperforming the state-of-the-art techniques and human reproduction.

    \item \textbf{Data.} Public released source code of {\tool} and the dataset of our experiments to facilitate the replication and extension of this study\footnote{Details are in our website: https://github.com/wuchiuwong/CrashTranslator\label{our_github}}.
\end{itemize}

\begin{figure*}[tb]
\setlength{\abovecaptionskip}{-0.1cm} 
\centering
\vspace{-0.05in}
\includegraphics[width=18.0cm]{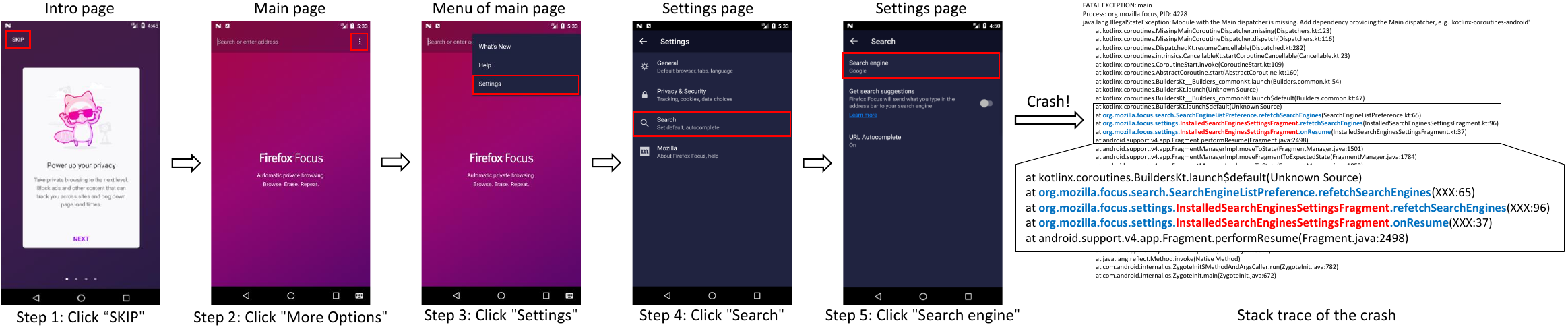}
\vspace{-0.05in}
\caption{Examples of crash reproducing}
\label{fig:all_example}
\vspace{-0.15in}
\end{figure*}

\section{Motivational Study}
\label{sec_movitation}

We conduct a motivational study to investigate whether it is common for stack-trace-only crash reports, their characteristics, and the challenges of reproducing these reports. 

In detail, we choose GitHub as the data source since it contains a large number of publicly available valid bug reports.
We use the web crawler provided by Wendland et al. \cite{wendland2021andror2} to automatically crawl the bug reports from Android projects and focus on the reports created from Jan. 2015 to May. 2022, resulting in 96,451 bug reports.
We then filter the bug reports involving application crashes with the keywords such as \textit{crash}, \textit{exception} following existing studies \cite{zhao2019ReCDroid, wendland2021andror2}. 
As a result, we acquire 10,843 Android crash reports for our motivational study.

\subsection{Is it Common for Stack-trace-only Crash?}
\begin{figure}[!htb]
\setlength{\abovecaptionskip}{0.1cm} 
\centering
\includegraphics[width=\columnwidth]{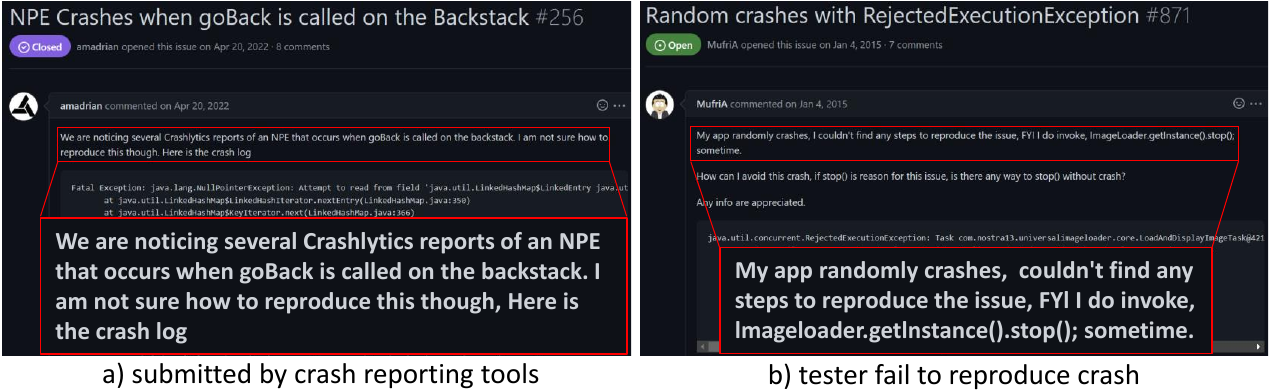}
\caption{Examples of stack-trace-only crash reports}
\label{fig:report_example}
\vspace{-0.1in}
\end{figure}

A well-formulated crash report usually contains the crash overview, textual-described reproducing steps, stack trace when the crash occurs, and visual recordings (screenshots/GIFs) about how the crash occurs. 
However, not all issue submitters would strictly follow the report template and provide all crash-related information when reporting the crash. 

We utilize keywords and heuristic pattern matching\textsuperscript{\ref{our_github}} to automatically examine whether the crash report contains reproducing steps and stack traces, following existing studies \cite{zhao2019ReCDroid, wendland2021andror2}.
Results reveal that 20.2\% (2,187 / 10,843) of crash reports contain only stack traces, i.e., stack-trace-only crash reports.
Such reports can be submitted by crash reporting tools such as Crashtics \cite{crashlytics}, which automatically collect crash logs and upload them to the issue server when the app crashes, as shown in Figure \ref{fig:report_example} (a).
Furthermore, commercial software can also have such auto-generated stack-trace-only crash reports \cite{ACRA, huynh2019ac3r}
, which further indicates the universality of such reports. 
Besides, these reports can also be submitted by the app users or developers who accidentally trigger the crash yet fail to figure out the reproducing steps, as shown in Figure \ref{fig:report_example} (b). 
This also implies the necessity for the automatic crash reproduction approach. 



\subsection{Is it Difficult to Handle Such Crash?}

We go a step further to investigate whether it is difficult to handle (e.g., reproduce, fix) such crashes. 
Since it can hardly obtain the time or effort for crash reproducing from GitHub, we turn to the commonly-used issue fixing duration, i.e., the duration between the issue creation time and the closing time \cite{de2022studying}, to indicate the difficulty of handling such crashes. 

For crash reports with reproducing steps, the average issue fixing duration is 57 days, while for stack-trace-only crash reports, such duration is increased to 72 days (26\% higher).
We assume that due to the insufficient information provided in the stack-trace-only crash report, developers need to spend extra effort in understanding the problem and reproducing the issue, and thus have a low willingness and take more time to fix the issue.
Therefore it would be highly expected to automate the reproduction of stack-trace-only crash reports to save the effort and facilitate follow-up issue fixing.


\subsection{Why is it Difficult?}

\textbf{Challenge 1: Lack of step-by-step guidance.}
Existing approaches would take the textual-described reproducing steps as input and conduct the bug reproduction guided by the steps \cite{zhao2019ReCDroid, zhao2022recdroid+, zhang2023automatically}.
However, for the stack-trace-only crash report, we cannot fetch the reproducing steps and thus lacks the step-by-step guidance to trigger the crash. 
By comparison, we can possibly derive the crash-occurring activity or fragment, i.e., the last interactive GUI page when the crash occurs, e.g., \textit{InstalledSearchEnginesSettingsFragment} as demonstrated in Figure \ref{fig:all_example}.
But the automatic approach still needs to speculate the exploration steps for navigating to the crash-occurring page. 
In our experimental dataset (shown in Section \ref{subsec_expdata}), there can be 1 to 8 exploration steps for reaching the crash-occurring GUI page, which can be quite inefficient if explored randomly. 

\textbf{Challenge 2: Need specific interactions even reaching the last GUI page.}
The second challenge is that even if the crash-occurring GUI page is reached, it may still need certain interactions with the app to finally trigger the crash. 
As shown in Figure \ref{fig:all_example}, to trigger the crash, after reaching the GUI page, one still needs to click \textit{Search engine} in the crash-occurring GUI page at step 5. 
In our experimental dataset (shown in Section \ref{subsec_expdata}), there can be an average of 6.6 interactive widgets, which further complicates the reproduction of stack-trace-only crashes.  

\subsection{Are There any Clues for Reproduction?}
While facing the above challenges, stack traces do provide clues for automated reproduction.
Specifically, as mentioned in challenge 1, we can derive the crash-occurring GUI page, which offers the target for the planned exploration. 
Besides, when conducting the operations in the crash-occurring GUI page, there can be corresponding invocations of the programming code, and the stack trace would output the involved APIs, e.g., \textit{refetchSearchEngines} and \textit{onResume} in Figure \ref{fig:all_example}.

To summarize, the stack trace offers two clues, i.e., the crash-occurring GUI page and the involved APIs when interacting with the app at the crash-occurring page.
We can utilize the first clue for predicting the exploration steps for navigating to the crash-occurring GUI page; and then use the second clue to find the widget by interacting with which the crash can finally be triggered. 

\mybox{
\textbf{Summary:} Our analysis on 10,843 crash reports of Android apps from GitHub shows that 20.2\% of crash reports only contain the stack trace, and these stack-trace-only reports consume 26\% more time for issue fixing. 
This can be because they lack step-by-step guidance for planning the reproduction. 
Our findings confirm the necessity and challenges of the crash reproduction directly from the stack trace. 
We also observe two clues to motivate our approach development for automated crash reproduction. 
}
\vspace{-0.05in}

\begin{figure*}[tb]
\setlength{\abovecaptionskip}{-0.1cm} 
\centering
\vspace{-0.05in}
\includegraphics[width=18.0cm]{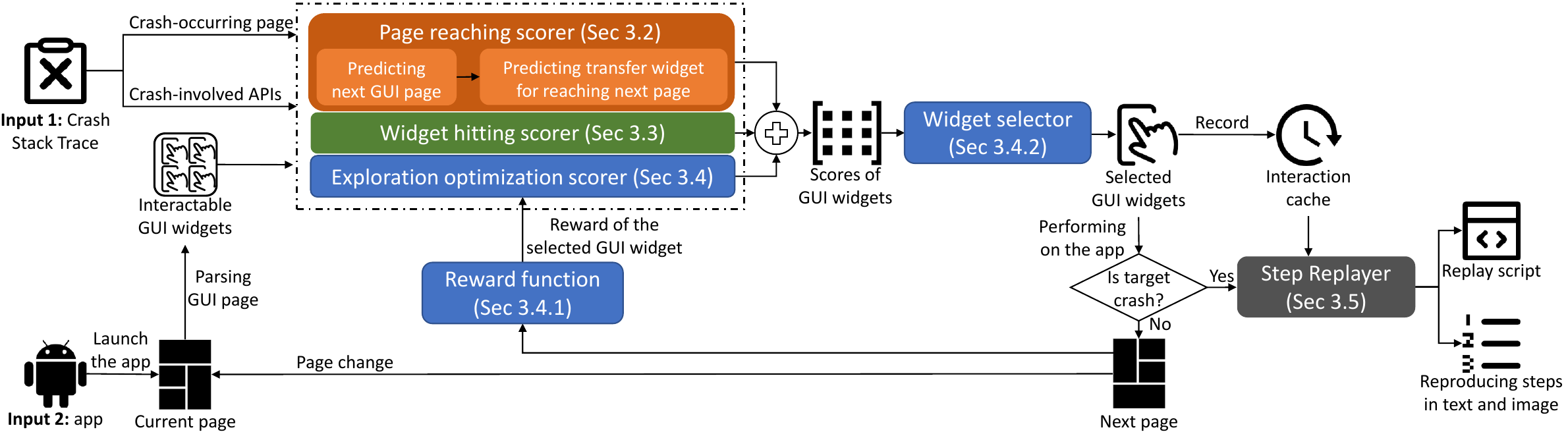}
\vspace{-0.05in}
\caption{Overview of {\tool}. \textit{Page reaching (Sec 3.2)} means arriving at the last GUI page when the crash occurs, and \textit{widget hitting (Sec 3.3)} means clicking the correct widget in the last GUI page for triggering the crash.}
\label{fig:overview}
\vspace{-0.15in}
\end{figure*}

\begin{figure}[thb]
\setlength{\abovecaptionskip}{0.05cm} 
\centering
\includegraphics[width=\columnwidth]{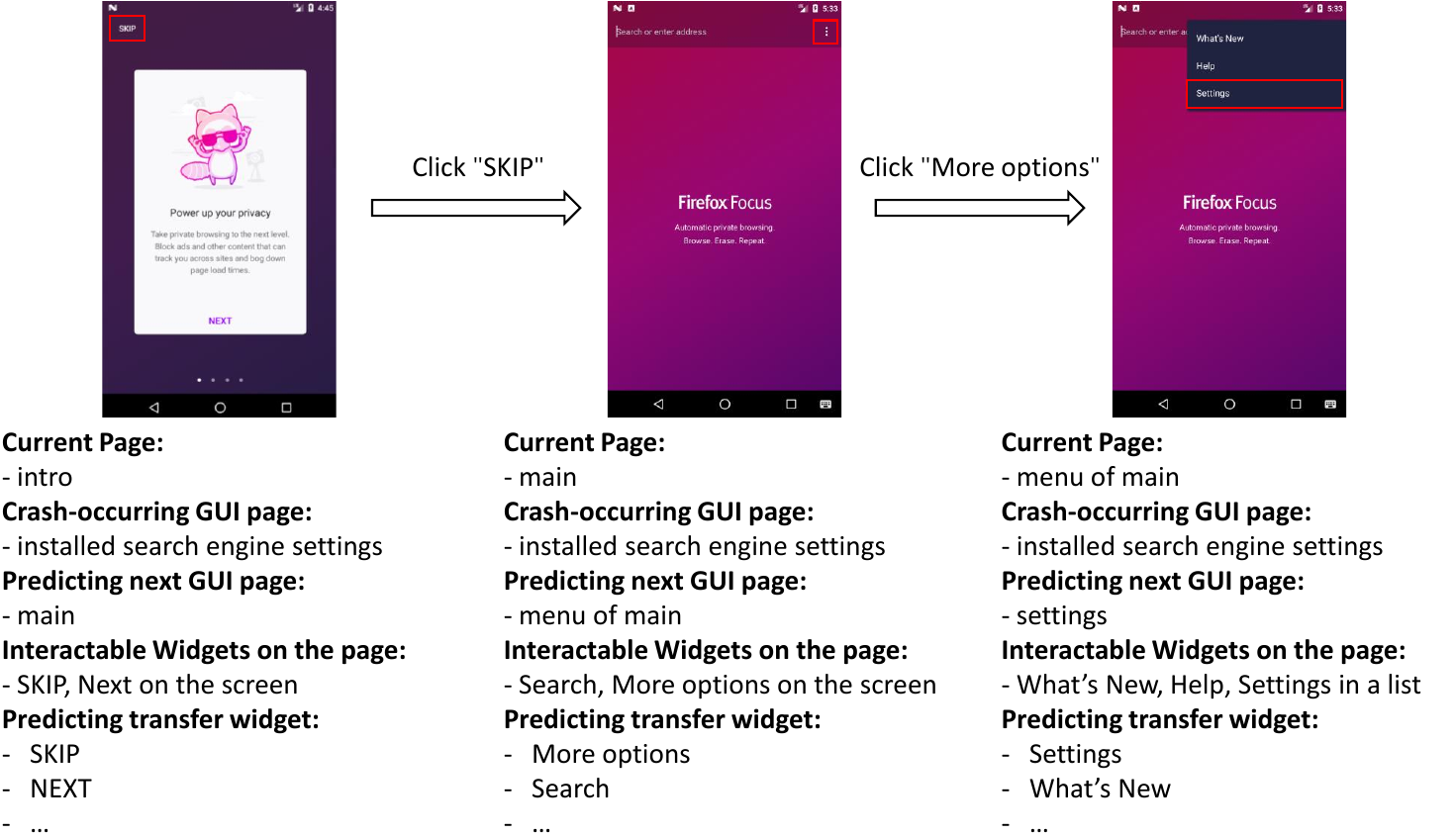}
\caption{Example of navigating from the app entry page to the crash-occurring page}
\label{fig:navigate_example}
\vspace{-0.1in}
\end{figure}

\section{APPROACH}
\label{sec_approach}
Motivated by the above findings, we propose an automated approach named \textit{CrashTranslator} to reproduce crash reports directly from the stack trace of mobile apps. 
It accomplishes this by leveraging a pre-trained Large Language Model (LLM) to predict the exploration steps for triggering the crash, and designing a reinforcement learning based technique to mitigate the inaccurate prediction and guide the search holistically. 

As demonstrated in Figure \ref{fig:overview}, given a stack-trace-only crash report, {\tool} first derives the crash-occurring GUI page and crash-involved APIs for triggering the crash. 
It designs three scorers to assign the exploration priority for each interactable GUI widget on the current page, i.e., 1) \textbf{Page reaching scorer} leverages the LLM to choose the next GUI page for reaching the crash-occurring GUI page, and the widget for transferring to the next page (Section~\ref{sec_approach_crash_page}); 2) \textbf{Widget hitting scorer} utilizes a heuristic method to find the crash-triggering widget by matching the crash-involved APIs (Section~\ref{sec_approach_crash_event}); and 3) \textbf{Exploration optimization scorer} assigns scores based on the previous interaction records of reinforcement learning technique, which aims at bridging the gap by inaccurate prediction of the first two scorers and plans the exploration holistically (Section~\ref{sec_approach_explore}).
{\tool} selects the GUI widget based on the three scorers and repeats the process iteratively until the target crash is triggered.  
It finally generates a replay script (for direct replay) and textual-described reproducing steps with step-by-step image instructions (for facilitating understanding).

\subsection{Preprocessing}
\label{sec_approach_preprocess}

We first conduct preprocessing for the stack trace and the target app, to prepare the information for reproducing the crash. 
Specifically, three types of information will be used in the crash reproduction.

\textbf{App's package name and the names of all activities in the app.} We first decompile the target app and get its configuration file (\textit{AndroidManifest.xml}) which records the package name of the app and the names of all activities it contains. 
In this paper, we consider the activity name as the GUI page name. 

\textbf{Crash-involved APIs.} 
We extract the lines that contain the app's package name from the stack trace, which indicate the crash-involved APIs in the app, e.g., \textit{org.mozilla.focus.search.SearchEngine-ListPreference.refetchSearchEngines} in blue color in Figure \ref{fig:all_example}.


\textbf{Crash-occurring page.} 
From the crash-involved APIs extracted in the second step, we then check whether the terms coincide with the app's activity name (extracted in the first step).
If so, we treat the activity as the crash-occurring page; otherwise, the crash may occur in a fragment, and we simply extract the name with the keywords \textit{Fragment} from the stack trace, e.g., \textit{InstalledSearchEnginesSettingsFragment} in red color in Figure \ref{fig:all_example}.

For better understanding and reducing noise, we further tokenize the extracted page names (activity names), crash-involved APIs, and the crash-occurring page by the underscore and Camel Case, and remove the stop words, for follow-up usage.





\subsection{Page Reaching Scorer}
\label{sec_approach_crash_page}
\begin{table*}[!t]

\setlength{\abovecaptionskip}{0.1cm}
\setlength{\belowcaptionskip}{0cm}
\caption{The example of prompt generation rules}
\centering
\scalebox{0.97}{
\begin{tabular}{p{0.4cm}p{1.5cm}p{2.1cm}p{2.1cm}p{2.1cm}p{2.1cm}p{2.1cm}p{2.1cm}p{2cm}}
\toprule
\multicolumn{9}{c}{\textbf{Input}} \\
\toprule
\multicolumn{1}{c|}{\textbf{ID}} & \multicolumn{1}{c}{\textbf{Attribute}} & \multicolumn{4}{|c|}{\textbf{Description}} & \multicolumn{3}{c}{\textbf{Examples}} \\
\midrule
\rowcolor{orange!10} \multicolumn{1}{l|}{I1} & PageNames & \multicolumn{4}{|p{8.4cm}|}{List of names of all activities in the app, extracted from AndroidManifest.xml file} & \multicolumn{3}{p{6.6cm}}{PageNames = {[}"intro", "main", "setting", …{]}} \\
\rowcolor{cyan!10} \multicolumn{1}{l|}{I2} & \hlc{CurrentPage} & \multicolumn{4}{|p{8.4cm}|}{The activity name of the current page} & \multicolumn{3}{p{6.6cm}}{CurrentPage = menu of main} \\
\rowcolor{purple!10} \multicolumn{1}{l|}{I3} & \hlp{CrashPage} & \multicolumn{4}{|p{8.4cm}|}{The name of the crash-occurring page, obtained from stack trace} & \multicolumn{3}{p{6.6cm}}{CrashPage = installed search engines settings} \\
\rowcolor{green!10} \multicolumn{1}{l|}{I4} & \multicolumn{1}{p{1.5cm}}{\hlg{Interactable-} \hlg{Widgets}} & \multicolumn{4}{|p{8.4cm}|}{List of names for all interactable GUI widgets on the current page, obtained from parsing the view hierarchy of the current page} & \multicolumn{3}{p{6.6cm}}{InteractableWidgets = {[}"what’s new", "help", "settings", …{]}} \\
\bottomrule
\multicolumn{9}{c}{\textbf{Prompt}} \\
\toprule
\multicolumn{1}{c|}{\textbf{ID}} & \multicolumn{1}{c}{\textbf{Task}} & \multicolumn{3}{|p{6.3cm}<{\centering}|}{\textbf{Prompt}} & \multicolumn{3}{|p{6.3cm}<{\centering}|}{\textbf{Instantiation}} & \multicolumn{1}{c}{\textbf{Prediction}} \\
\midrule
\multicolumn{1}{l|}{P1} & Predicting next GUI page & \multicolumn{3}{|p{6.3cm}|}{There are \hlo{\textless{}\textit{\#NumOfPageNames}\textgreater} \space pages in the app, named: \hlo{\textless{}\textit{PageNames}\textgreater} . I want to go from the \hlc{\textless{}\textit{CurrentPage}\textgreater} \space page to the \hlp{\textless{}\textit{CrashPage}\textgreater} \space page. What is the next page?} & \multicolumn{3}{|p{6.3cm}|}{There are \hlo{8} pages in the app, named: \hlo{intro, main, setting, ...} . I want to go from the \hlc{menu of main page} to the \hlp{installed search engines settings} page. What is the next page?} & \hlgg{\textless{}\textit{NextPage}\textgreater} = setting \\
\multicolumn{1}{l|}{P2} & Predicting transfer widget for reaching next page & \multicolumn{3}{|p{6.3cm}|}{There are \hlo{\textless{}\textit{\#NumOfPageNames}\textgreater} \space pages in the app, named: \hlo{\textless{}\textit{PageNames}\textgreater} . I want to go from the \hlc{\textless{}\textit{CurrentPage}\textgreater} \space page to the \hlp{\textless{}\textit{CrashPage}\textgreater} \space page. The next page may be the \hlgg{\textless{}\textit{NextPage}\textgreater} \space page. Here are widgets I can click:\hlg{\textless{}\textit{InteractableWidgets}\textgreater{}}. What should I click?} & \multicolumn{3}{|p{6.3cm}|}{There are \hlo{8} pages in the app, named: \hlo{intro, main, setting, ...} . I want to go from the \hlc{menu of main} page to the \hlp{installed search engines settings} page. The next page may be the \hlgg{setting} page. Here are widgets I can click: \hlg{what’s new, help, settings in a list, ...}. What should I click?} & \textless{}\textit{TransferWidget} \textgreater ~= settings \\
\bottomrule
\end{tabular}
}
\vspace{-0.08in}
\label{tab:prompt_example}
\end{table*}


To reach the crash-occurring GUI page, intuitively, we can interact with the GUI widgets that share similar names with the crash-occurring page, e.g., click \textit{Search} in step 4 can reach the crash-occurring \textit{installed search engines settings fragment} as shown in Figure 
\ref{fig:all_example}.
However, there can be a long sequence of exploration steps for reaching the crash-occurring page, and the widgets in the prior and middle 
part of the sequence tend to be irrelevant to the crash-occurring page, e.g., \textit{More options} in step 2 looks totally different from the crash-occurring page as shown in Figure 
\ref{fig:all_example}.
To tackle this, we leverage the LLM to predict the exploration steps for reaching the crash-occurring page iteratively. 
Specifically, as shown in Figure \ref{fig:navigate_example}, we first ask the LLM to predict the next page for reaching the crash-occurring page; then ask the LLM to choose the widget for transferring to the predicted next page; and iterate the process.
This step would assign scores for the widgets, indicating their probabilities of reaching the crash-occurring GUI page, i.e., priorities of being chosen.


\subsubsection{\textbf{Predicting Next Page for Reaching Crash-occurring Page}}

We provide the LLM with all app’s pages and the current GUI page, and ask the LLM to speculate the next page to reach the crash-occurring page, by which the whole exploration sequence is generated iteratively. 



\textbf{Input.}
There are three inputs, i.e., I1-I3 as shown in Table \ref{tab:prompt_example}. 
1) PageNames, i.e., the name of all activities, are extracted from the \textit{AndroidManifest.xml} as described in Section \ref{sec_approach_preprocess};
2) CurrentPage is the activity name of the current GUI page during the iterative process.
To distinguish between pages with the same activity name but with different widgets on them (e.g., the pages in steps 2 and 3 in Figure \ref{fig:all_example}), we divide the pages into three types: menu, dialog, and general pages. For menu or dialog page, we name the CurrentPage by "\textit{page\_type of activity\_name}" (e.g., \textit{menu of main} in step 3 in Figure \ref{fig:all_example}). 
3) CrashPage, i.e., the name of the crash-occurring page, is extracted from the crash stack trace as described in Section \ref{sec_approach_preprocess};

\textbf{Prompt generation.} 
To design the prompt, we follow the regular prompt template \cite{chen2022knowprompt, Cantino201Prompt, gu2021ppt}, and each of the three authors is asked to write the prompt sentence for this task with 10 trial apps, and conduct a discussion to derive the final prompt pattern, i.e., P1 as shown in Table \ref{tab:prompt_example}.
Take the last step in Figure \ref{fig:navigate_example} as an example, we first tell the LLM what pages are contained in the app (\textit{There are 8 pages in the app, named: intro, main, setting, ...}), then tell the LLM what we want to do with the information about the current page and crash-occurring page (\textit{I want to go from the menu of main page to the installed search engines settings page}), and finally ask what the next page should be (\textit{What is the next page?}). 
The LLM will predict the name of the next page (\textit{setting}).




\textbf{Fine-tuning.}
To achieve better performance, we build a fine-tuning dataset and fine-tune the LLM for learning the transition relations between app pages. 
Specifically, we collect 1,000 apps of different categories from F-Droid \cite{fdroid} and extract their activity transition graph (ATG) with Gator \cite{gator}, one of the state-of-the-art static analysis tools.
We then utilize these transition relations as fine-tuning data for model fine-tuning. 
Although the static analysis tool like Gator cannot obtain the full ATG, which is why we do not directly use it for path planning, yet through inputting the incomplete ATG from different apps, the LLM has the potential to combine the diversified viewpoints together and speculate the desired transitions. 
Note that the fine-tuning process is a one-time requirement and does not necessitate individual fine-tuning for each app.


\subsubsection{\textbf{Predicting Transfer Widget for Reaching Next Page}}


After knowing the next page, we need to know how to reach there, e.g., which button to click. 
We provide the LLM with all interactable GUI widgets on the current page, then ask the LLM to choose the widget for transferring to the next page.


\textbf{Input.} 
Besides the three inputs used in the previous section, this step needs the fourth input, i.e., InteractableWidgets (I4 as shown in Table \ref{tab:prompt_example}).
It is the list of names for all widgets which can be interacted with on the current GUI page, and we obtain it from the view hierarchy file of the current page. 
Specifically, we first filter the interactive widgets from all the widgets based on whether the \textit{clickable} or \textit{long-clickable} property is true.
Then, we leverage heuristic rules to extract a representative name for each widget following existing studies \cite{zhao2019ReCDroid, mariani2021semantic, mehralian2021data}.
In detail, for text-like widgets (e.g., \textit{Buttons}, \textit{TextView}, etc.), we obtain their textual attributes (e.g., \textit{text}, \textit{content-description}, \textit{resource-id}) and utilize the first non-empty one. 
For icon-like widgets (e.g., \textit{ImageButton}, \textit{ImageView}, etc.), we extract their name from their contextual text information (e.g., \textit{nearby text}, \textit{sibling text}, \textit{child text}) and use the first non-empty one. 
Finally, we tokenize the extracted widget name by the underscore and Camel Case, and group widgets according to their container widgets to express the current page's layout, e.g., \textit{What's New, Help, Settings in a list} in step 3 of Figure \ref{fig:navigate_example}.

\textbf{Prompt generation.} Following the same procedure as the previous section, we come out with the prompt pattern, i.e., P2 in Table \ref{tab:prompt_example}.
Take the last step in Figure \ref{fig:navigate_example} as an example. 
Like the prompt in the previous section, we first tell the LLM what pages are contained in the app and our ultimate goal; we then tell it the predicted next page (\textit{The next page may be the setting page}), provide the names of all widgets on the current page (\textit{Here are widgets I can click: what’s new, help, settings in a list, ...}), and finally ask LLM to choose the optimal transferring widget (\textit{What should I click?}).
The LLM will output a widget which it thinks could lead to the target page (\textit{settings}).


Since the LLM might not always output the correct transfer widget, we would let it provide a ranked list of candidate widgets (i.e., top 5), and {\tool} would consider these ranked widgets during the exploration optimization in Section \ref{sec_approach_explore}.
To realize this, we repeat the widget prediction by utilizing a new prompt in which we remove the interactable widgets that have already been predicted, and let the LLM provide a new answer.
This process is repeated five times, and we obtain five distinct widgets in order. 
We then assign numerical scores to each widget in the manner of $1/(rank+2)$ (e.g., Top 1 scores 0.33, Top 2 scores 0.25).


\textbf{Fine-tuning.} 
Similar to the prior section, we fine-tune the LLM for better performance. 
We use the commonly-used RICO dataset \cite{deka2017rico}, which contains plenty of the targeted GUI page and transfer widget for reaching it. We randomly sample 1,000 such data pairs (involving 629 apps) as the fine-tuning dataset. Note that the fine-tuning process here also only needs to be conducted once.

\subsection{Widget Hitting Scorer}
\label{sec_approach_crash_event}

Some crashes can be triggered as soon as arriving at the crash-occurring page, yet in most cases, specific events or event combinations are needed to perform on the crash-occurring page to finally trigger the crash, e.g., although we reach the \textit{installed search engine settings} page at step 4 in Figure \ref{fig:all_example}, the crash has not yet been triggered until we click the \textit{Search engine} at step 5.
We utilize the crash-involved APIs to infer the crash-triggering widgets. Take Figure 1 as an example, the stack trace involves API of \textit{refetchSearchEngines}, from which we can infer that clicking the \textit{Search engine} on the crash-occurring page would likely trigger the crash. 



To automatically find these crash-triggering widgets, we propose a lightweight matching method 
between widgets and crash-involved APIs extracted from the stack trace.
Specifically, we first tokenize the name of widgets and crash-involved APIs by the underscore and Camel Case. 
If there is an overlap at the token level between the name of a widget and an API, i.e. at least one token is the same after stemming or an abbreviation of the other, we assume the widget is a candidate crash-triggering widget and assign a score based on the percentage of overlapping tokens.

\subsection{Exploration Optimization Scorer}
\label{sec_approach_explore}

Ideally, the crash can be reproduced with the predicted transfer widget for arriving at the crash-occurring page (in Section \ref{sec_approach_crash_page}) and the crash-triggering widget after reaching the crash-occurring page (in Section \ref{sec_approach_crash_event}).
Yet, in practice, the prediction can be inaccurate, and there might be complex transitions in the app which the prediction model could not capture.
Therefore we design the exploration optimization scorer to help plan the exploration holistically and bridge the gap by inaccurate prediction of the first two scorers. 

We leverage Q-learning \cite{watkins1992q}, a reinforcement learning method, to help conduct the exploration. 
The basic idea is to maintain a Q-table, which stores all widgets’ value, to record the crash-related and exploration information and memorize the valuable widgets during trial-and-error exploration. 
When we first reach a new page, the value of each widget on the page is initialized to 0 and updated by reward or penalty after the widget’s interaction (details below). 

\subsubsection{\textbf{Formulating Exploration as MDP}}

In our approach, we define an instance of the Markov decision process (MDP) to describe the exploration process and adopt Q-learning to optimize the exploration. The MDP can be defined with a 4-tuple, $\langle\mathcal{S}$,$\mathcal{A}$,$\mathcal{P}$,$\mathcal{R}\rangle$, where $\mathcal{S}$ refers to the set of \textbf{states}, $\mathcal{A}$ refers to the set of \textbf{actions}, $\mathcal{P}$ refers to the \textbf{transition function}, and $\mathcal{R}$ refers to the \textbf{reward function}. In the context of crash reproduction, each page in the app represents an individual state. We extract interactable GUI widgets on each page, and interactions (click, long click or type text) with widgets constitute the action set of the state. When we perform an action $a_t$ (an interaction on a widget), the app's state will be changed from state $s_t$ to state $s_{t+1}$. We first record the transition $\langle s_t, a_t, s_{t+1} \rangle$ to the transition function $\mathcal{P}$, and then assign a reward $r_t$ based on the reward function $\mathcal{R}$.
The reward function $\mathcal{R}$ generates a value indicating the quality of a performed action, e.g., whether interacting with a certain widget relates to the crash. 
Our reward function evaluates an action (i.e., the corresponding widget) as a sum of the following three aspects.

\textbf{Crash triggering reward}. When an action involves the crash-related elements, i.e., reaching the crash-occurring page or triggering crash-involved APIs, the corresponding widget will receive a large positive reward since the exploration in these widgets has a larger possibility of triggering the crash. 
Next time when the approach reaches the page, it might choose the widget again, and the combination containing the widget might trigger the crash.

\textbf{New state reward}. When an action explores a new GUI page, the corresponding widget will receive a small positive reward to encourage exploiting new states, especially at the beginning of the exploration.
With the exploration going on, the new state reward of a widget can be balanced out by the duplicate state penalty. 

\textbf{Duplicate or failure state penalty}. When a widget transfers to a known page, it will receive a small penalty since it is less likely to trigger the crash.
Besides, when an action transfers to a page out of the app (e.g., opening a browser) or triggers a non-target crash, the corresponding widget will receive a large penalty. It will not be chosen again since it is impossible to trigger the crash.

Next, we will update the value of the widget interaction $Q(s_t, a_t)$ recorded in the Q-table with the Bellman function:
\begin{equation}
    Q(s_t, a_t) \leftarrow Q(s_t, a_t) + \alpha(r_t+\gamma Q^*(s_{t+1}, a_{t+1})-Q(s_t, a_t))
\end{equation}
where $\alpha$ refers to the learning rate and is set to 0.1, $\gamma$ refers to the discount factor and is set to 0.9, $Q^*(s_{t+1}, a_{t+1})$ refers to the maximum value of all actions in state $s_{t+1}$.

\subsubsection{\textbf{Widget selector}}
\label{subsubsec_selector}

To efficiently focus on the potentially correct planning and also break out the local optimum, we leverage the $\epsilon$-greedy policy \cite{thrun2000reinforcement} to select the widget to be performed following existing studies \cite{pan2020reinforcement, zhang2023automatically}.
Specifically, for each widget, we sum the assigned score from the three scorers (i.e., page reaching scorer, widget hitting scorer, exploration optimization scorer); then choose the widget with the highest sum score with a high probability $1-\epsilon$, or randomly select other widgets with a low probability $\epsilon$. 
In practice, $\epsilon$ is initially set as a small number close to 0 to enable {\tool} to focus on the predicted optimal widget (i.e., the one with the highest sum score). 
During exploration, the optimal widget may be wrongly predicted and lead the exploration stuck in a repetitive exploration between several pages; at this point, 
$\epsilon$ will be changed to a big number close to 1, leading {\tool} to select the non-optimal widget to break the local optimum.


\subsection{Step Replayer}
\label{sec_approach_gen_step}
During crash reproducing, {\tool} uses an interaction cache to record all interactions with widgets when the app is launched (and the cache will be cleared after the app restart). After the target crash is triggered, the interaction recorded in the cache is crash-reproducing operations from app launch to crash. 
However, the interactions recorded in the cache may not be the most straightforward way to trigger the crash and might contain some redundant interactions.
To make the generated reproducing steps more concise, we automatically eliminate the redundant interactions that lead to repeated or looped transitions.
Finally, we convert the concise interaction history into an auto-replay script and human-readable reproducing steps, i.e., ``image + text'' reproducing steps shown in Figure \ref{fig:all_example}. For each interaction, we highlight the widget to be interacted with on the page’s screenshot by a red box and generate the textual reproducing step in the form of ``event type + widget name''.

\subsection{Implementation}

We implement our approach in Python and extend functionalities from the following libraries: Appium \cite{appium} to interact with Android apps and obtain the view hierarchy of the current page; 
NLTK \cite{nltk} to stem word, which is used in the widget hitting scorer (Section \ref{sec_approach_crash_event}); 
Ella \cite{ella} to check whether crash-involved APIs are triggered, which is used in the exploration optimization scorer (Section \ref{sec_approach_explore}); 
OpenCV \cite{opencv} to mark widgets that need to be interacted with on screenshots, which is used in generating reproducing steps (Section \ref{sec_approach_gen_step}).
We run {\tool} and perform experiments on a physical x86 Ubuntu 20.04 machine with Android emulators (Android 4.4-7.0).

For the LLM leveraged in the page reaching scorer (Section \ref{sec_approach_crash_page}), we adopt the pre-trained GPT-3 \cite{brown2020language} model from OpenAI\footnote{https://platform.openai.com/docs/models/gpt-3}.
We choose the Curie model as the base model and fine-tune the model through official APIs as described in Section \ref{sec_approach_crash_page}.
Developers only need to set up their OpenAI account, complete the fine-tuning process according to the instructions provided on our website\textsuperscript{\ref{our_github}}, and subsequently utilize our tool to automatically. On average, reproducing one crash requires sending approximately 42.6 prompts (4939.7 tokens), with an estimated cost of around 0.01 USD.

\section{EXPERIMENT Design}
\label{sec_exp}

To evaluate {\tool}, we consider the following three research questions:

\textbf{RQ1}: How effective and efficient is {\tool} in reproducing crashes from stack trace?

\textbf{RQ2}: What is the contribution of the designed scorers in {\tool} for reproduction?

\textbf{RQ3}: Can the reproducing steps generated by {\tool} help developers to reproduce crashes?

\subsection{Experimental Dataset}
\label{subsec_expdata}

In this work, We collect 75 crash reports involving 58 apps from three sources for evaluation, i.e., ReCDroid's dataset \cite{zhao2019ReCDroid}, AndroR2 dataset \cite{wendland2021andror2}, and GitHub. 
ReCDroid is an approach for crash replay based on the textual-described reproducing steps, and we utilize all 33 crash reports in its replicate package. 
AndroR2 is a dataset of manually-reproduced bug reports for Android apps, and we use all its 22 crash reports. 
Other reports, e.g., display bug reports, are out of the scope of this study.
Note that the above 55 (33+22) reports do not necessarily contain stack traces. 
For those reports without a stack trace, we manually reproduce the crash following the reproducing steps and then extract the stack trace from the log. 


We also collect a third dataset from GitHub to further prove the effectiveness of {\tool}. In detail, we first crawl and filter 3,566 crash reports with the stack trace (may also contain reproducing steps) from GitHub as described in Section \ref{sec_movitation}, and randomly sample 300 crash reports for manual checking to retrieve the reproducible ones. 
It is performed independently by three graduate students with 2-4 years of Android development experience, and each report is manually reproduced by two of them. We exclude those that cannot be reproduced (e.g., lack of apks, failed-to-compile apps, environment issues) or require special conditions (e.g., account, hardware). This results in 20 crash reports involving 15 apps, and we refer to this dataset as {\tool}’s dataset.

Note that the crash reports used in the experiments may contain reproducing steps or screenshots, but {\tool} will not use such information and only uses the stack trace to reproduce crashes. 
Due to space limitations, the details of the dataset can be viewed on our website\textsuperscript{\ref{our_github}}.

\subsection{Baselines}
To the best of our knowledge, {\tool} is the first work to reproduce crashes directly from the stack trace. 
Existing studies which reproduce crashes from the natural language described reproducing steps, e.g., RecDroid \cite{zhao2019ReCDroid} and ReproBot \cite{zhang2023automatically}, can not work for our task. 
Nevertheless, we forcefully apply ReCDroid as a baseline for our task and donate as \textbf{ReCDroid$_{ST}$}. Specifically, We provide the crash stack trace as its input rather than the reproducing steps, irrespective of whether ReCDroid comprehends the stack trace or not.

In addition, there are automated GUI testing approaches \cite{lv2022fastbot2, su2017guided, wang2020combodroid, dong2020time, monkey, li2019humanoid, gu2019practical, pan2020reinforcement, mao2016sapienz} which also explore the app and try to reveal the crashes; hence we utilize these approaches as the baselines to better prove our effectiveness. 
We choose the following four state-of-the-art approaches from different categories, i.e., Monkey, Humanoid, APE, and Q-testing:

\textbf{Monkey} \cite{monkey} is a widely-used random-based GUI testing tool that tests the target app with purely random sequences of GUI events or system events. The advantages of Monkey are its ability to perform lots of GUI events quickly and its good compatibility.

\textbf{Humanoid} \cite{li2019humanoid} is a novel deep learning-based GUI testing tool. It trains a deep neural network model from a large-scale crowd-sourced human interactions dataset to predict which GUI widgets on the current page are more likely to be interacted with by testers. 

\textbf{Ape} \cite{gu2019practical} is one of the state-of-the-art model-based GUI testing tools. It models the app's behavior by building a finite state machine dynamically. The advantage of Ape is its ability to balance the size and precision of the modelling by using dynamic GUI abstraction.

\textbf{Q-testing} \cite{pan2020reinforcement} uses reinforcement learning to guide testing toward new pages to find crashes. It rewards GUI events that reach a new page and penalizes events that transfer to an explored page.

\subsection{Experimental Setup and Evaluation Metrics}
\label{subsec_expsetup}
For RQ1, we verify the effectiveness and efficiency of {\tool} in two aspects: (1) the percentage of reports that can be successfully reproduced in a given time (denoted as \textit{success rate}). We set one hour following the existing study \cite{pan2020reinforcement}. (2) The time required for successful reproduction (denoted as \textit{reproducing time}). 
To mitigate the bias from randomness, we run our approach and the baselines three times and record the average reproducing time.

For RQ2, to investigate the contribution of the designed scorers, we would remove each of them and evaluate the performance for crash reproduction. 
Note that since the exploration optimization scorer is responsible for the exploration, if removing it, the exploration might be stuck in a local dilemma and could not finish the reproduction.
Therefore, we only evaluate the contribution of the other two scorers. 
In detail, we create two variants of {\tool}, i.e., 1) \textbf{CT$_{wp}$}, the variant \textbf{w}ithout the \textbf{p}age reaching scorer described in Section \ref{sec_approach_crash_page}. 2) \textbf{CT$_{ww}$}, the variant \textbf{w}ithout the \textbf{w}idget hitting scorer described in Section \ref{sec_approach_crash_event}. 
The relative contribution of each scorer is measured by comparing each variant with the original approach in terms of success rate and reproducing time, which is also based on the average of three runs. 

For RQ3, we verify the usefulness of reproducing steps generated by {\tool}. We invite 14 postgraduate students to participate in this experiment. All of them have experience in mobile application testing, 8 are Android developers with at least 3 years of development experience, and 7 work in the crowdtesting platform.
For the 46 crash reports that can be reproduced by {\tool}, we ask participants to reproduce the crash manually based on the stack trace or reproducing steps (generated by \tool).
Specifically, each participant is assigned 20 different reports, 10 of which only contain the stack trace, while the other 10 only contain the reproducing steps, thus ensuring that each report is reproduced by 3 participants based on the stack trace and 3 others based on steps. If a participant can reproduce the crash within 30 minutes following the existing study \cite{zhao2019ReCDroid}, we record the reproducing time; otherwise, we mark it as failing to reproduce. Finally, we compare the success rate and reproducing time based on the stack trace and {\tool}-generated reproducing steps.


\section{Results and Analysis}
\subsection{RQ1: Effectiveness and Efficiency}
\label{sec_result_rq1}

Table \ref{tab:expe_succ} shows the success rate of reproducing crash reports from three datasets. Overall, {\tool} can reproduce 61.3\% of them (46 out of 75), outperforming the baselines by a large margin, i.e., 
171\% (61.3\% vs. 22.6\%) higher than ReCDroid$_{ST}$,
206\% (61.3\% vs. 20\%) higher than Monkey, 142\% (61.3\% vs. 25.3\%) higher than Humanoid, 109\% (61.3\% vs. 29.3\%) higher than Ape, 206\% (61.3\% vs. 20\%) higher than Q-Testing. This shows that our tool can effectively reproduce crashes based on the corresponding stack trace. Specifically, {\tool} successfully reproduces 28 (84.8\%) reports on the ReCDroid's dataset, which outperforms ReCDroid$_{ST}$ and other automated GUI testing baselines (30.3\%-51.5\%). While on the AndroR2 dataset and the {\tool}’s dataset, {\tool} can achieve a success rate of 40.9\%-45\%. In contrast, ReCDroid$_{ST}$ and other automated GUI testing baselines can only successfully reproduce a small portion of reports, with a success rate of 5\% to 13.6\%. 
The difference in the success rate of the three datasets is might because that crash reports in the ReCDroid's dataset tend to involve fewer exploration steps for reproduction, while the other two datasets require a longer exploration sequence for reproducing the crash. 

For the 29 reports that {\tool} fails to reproduce, there are three main reasons for hindering the reproduction: 
1) Some apps (3 from the ReCDroid's dataset, 1 from the AndroR2 dataset) could not run in our environment, e.g., the server is down, incompatibility with our emulator.
2) {\tool} does not cover all the interactive GUI actions. It already supports such GUI actions as tapping, long pressing and typing, and rotating the screen like existing studies \cite{zhao2019ReCDroid, zhao2022recdroid+}, yet some crashes require other types of actions to trigger (e.g. \textit{C-4 Alarmio-47} requires scrolling up and down the screen). 3) Other unsolved technical challenges, which are further discussed in the discussion (Section \ref{subsec_dislim}).

\begin{table}[!t]
\centering
\vspace{0.05in}
\setlength{\abovecaptionskip}{0cm}
\setlength{\belowcaptionskip}{-0.2cm}
\caption{Reproduction success rate (Effectiveness). \textbf{CT}, \textbf{R}, \textbf{M}, \textbf{H}, \textbf{A}, \textbf{Q} in the table header refers to {\tool}, ReCDroid$_{ST}$, Monkey, Humanoid, Ape and Q-testing respectively.}
\scalebox{0.59}{
\begin{tabular}{r|rrrrrr}
\toprule
\textbf{\#   Dataset} & \multicolumn{1}{l}{\textbf{CT}} & \multicolumn{1}{l}{\textbf{R}} & \multicolumn{1}{l}{\textbf{M}} & 
\multicolumn{1}{l}{\textbf{H}} & \multicolumn{1}{l}{\textbf{A}} & \multicolumn{1}{l}{\textbf{Q}} \\
\hline
\textbf{ReCDroid's Dataset (33)} & \textbf{27 (81.8\%)} & 14 (42.4\%) & 11 (33.3\%) & 15 (45.5\%) & 17 (51.5\%) & 10 (30.3\%) \\
\textbf{AndroR2 Dataset (22)} & \textbf{10 (45.5\%)} & 2 (9.1\%) & 0 (0\%) & 2 (9.1\%) & 3 (13.6\%) & 3 (13.6\%) \\
\textbf{{\tool}'s Dataset (20)} & \textbf{9 (45\%)} & 1 (5\%) & 2 (10\%) & 2 (10\%) & 2 (10\%) & 2 (10\%) \\ \hline
\textbf{total (75)} & \textbf{46 (61.3\%)} & 17 (22.7\%) & 13 (17.3\%) & 19 (25.3\%) & 22 (29.3\%) & 15 (20\%)                        \\ \bottomrule
\end{tabular}
}
\vspace{-0.1in}
\label{tab:expe_succ}
\end{table}
\begin{table}[!t]
\centering
\vspace{0.05in}
\setlength{\abovecaptionskip}{0cm}
\setlength{\belowcaptionskip}{-0.2cm}
\caption{Reproducing time on successfully reproduced reports (Efficiency)}

\scalebox{0.78}{
\begin{tabular}{r|rrrrrr}
\toprule
\multicolumn{1}{l|}{\textbf{Reproducing   Time (s)}} & \multicolumn{1}{l}{\textbf{Average}} & \multicolumn{1}{l}{\textbf{Min}} & \multicolumn{1}{l}{\textbf{Q$_1$}} & \multicolumn{1}{l}{\textbf{Median}} & \multicolumn{1}{l}{\textbf{Q$_3$}} & \multicolumn{1}{l}{\textbf{Max}} \\ \hline
ReCDroid$_{ST}$ (17 reports) & 988 & 10 & 94 & 996 & 1436 & 2719 \\
Monkey (15 reports) & 669 & 5 & 48 & 152 & 985 & 2525 \\
Humanoid (19 reports) & 532 & 12 & 43 & 216 & 1094 & 1642 \\
Ape (22 reports) & 531 & 5 & 33 & 84 & 809 & 2166 \\
Q-testing (15 reports) & 368 & 6 & 30 & 118 & 548 & 1506 \\ \hline
\textbf{{\tool}} (46 reports) & 68.7 & 8 & 19 & 44 & 87 & 640 \\
\bottomrule
\end{tabular}
}
\vspace{-0.1in}
\label{tab:expr_time_stat}
\end{table}
\begin{table*}[!t]
\caption{Reproduction details on three datasets. \textbf{CT}, \textbf{R}, \textbf{M}, \textbf{H}, \textbf{A}, \textbf{Q} in the table header refers to {\tool}, ReCDroid$_{ST}$, Monkey, Humanoid, Ape and Q-testing respectively. \textbf{CT$_{wp}$} and \textbf{CT$_{ww}$} refer to variants of {\tool} without the page reaching scorer and without the widget hitting scorer, respectively. If the above approach successfully reproduces the crash, we record the reproduction time (in seconds) in the table, and if it fails, we record $\times$. Columns P$_{trace}$ and P$_{step}$ show the average time for participants to manually reproduce the crash based on the stack trace or reproducing steps generated by {\tool}, respectively, and the number in parentheses is the number of participants who reproduce crash successfully (out of 3). Note that in order to save space, reports that cannot be reproduced by either approach are omitted.}
\vspace{-0.1in}
\centering
\setlength{\tabcolsep}{4pt}
\scalebox{0.7}{
\begin{tabular}{ll|rrrrrr|rr|rr|ll|rrrrrr|rr|rr}
\toprule
 &  & \multicolumn{6}{c|}{\textbf{RQ 1}} & \multicolumn{2}{c|}{\textbf{RQ 2}} & \multicolumn{2}{c|}{\textbf{RQ 3}} &  &  & \multicolumn{6}{c|}{\textbf{RQ 1}} & \multicolumn{2}{c|}{\textbf{RQ 2}} & \multicolumn{2}{c}{\textbf{RQ 3}} \\ \hline
\multicolumn{1}{l|}{\textbf{ID}} & \textbf{\#Crash Reports} & \multicolumn{1}{l}{\textbf{CT}} & \multicolumn{1}{l}{\textbf{R}} & \multicolumn{1}{l}{\textbf{M}} & \multicolumn{1}{l}{\textbf{H}} & \multicolumn{1}{l}{\textbf{A}} & \multicolumn{1}{l|}{\textbf{Q}} & \multicolumn{1}{l}{\textbf{CT$_{wp}$}} & \multicolumn{1}{l|}{\textbf{CT$_{ww}$}} & \multicolumn{1}{l}{\textbf{P$_{trace}$}} & \multicolumn{1}{l|}{\textbf{P$_{step}$}} & \multicolumn{1}{l|}{\textbf{ID}} & \textbf{\#Crash Reports} & \multicolumn{1}{l}{\textbf{CT}} & \multicolumn{1}{l}{\textbf{R}} & \multicolumn{1}{l}{\textbf{M}} & \multicolumn{1}{l}{\textbf{H}} & \multicolumn{1}{l}{\textbf{A}} & \multicolumn{1}{l|}{\textbf{Q}} & \multicolumn{1}{l}{\textbf{CT$_{wp}$}} & \multicolumn{1}{l|}{\textbf{CT$_{ww}$}} & \multicolumn{1}{l}{\textbf{P$_{trace}$}} & \multicolumn{1}{l}{\textbf{P$_{step}$}} \\ \hline
\multicolumn{1}{l|}{R-1} & NewsBlur-1053 & 32 & 176 & 2024 & $\times$ & 68 & $\times$ & 53 & 69 & 201 (3) & 69 (3) & \multicolumn{1}{l|}{R-25} & K-9Mail-2612 & 11 & $\times$ & $\times$ & $\times$ & $\times$ & $\times$ & 22 & $\times$ & 194 (3) & 65 (3) \\
\multicolumn{1}{l|}{R-2} & Markor-194 & 88 & $\times$ & $\times$ & $\times$ & 2166 & 1156 & 52 & 94 & 540 (1) & 120 (3) & \multicolumn{1}{l|}{R-26} & K-9Mail-2019 & 12 & $\times$ & $\times$ & $\times$ & $\times$ & $\times$ & 16 & 21 & 185 (2) & 68 (3) \\
\multicolumn{1}{l|}{R-3} & Birthdroid-13 & 112 & 1776 & $\times$ & $\times$ & 1564 & $\times$ & 100 & 127 & 230 (2) & 73 (3) & \multicolumn{1}{l|}{R-27} & TagMo-12 & 42 & $\times$ & 52 & 47 & 12 & $\times$ & 25 & 42 & 176 (2) & 43 (2) \\
\multicolumn{1}{l|}{R-4} & Car Report-43 & 46 & 2558 & $\times$ & $\times$ & $\times$ & $\times$ & 58 & 46 & 475 (2) & 92 (3) & \multicolumn{1}{l|}{R-28} & FlashCards-13 & 23 & $\times$ & $\times$ & $\times$ & $\times$ & $\times$ & 21 & 21 & 73 (2) & 18 (3) \\
\multicolumn{1}{l|}{R-5} & AnyMemo-18 & 19 & $\times$ & 125 & 644 & 111 & $\times$ & $\times$ & 23 & 63 (3) & 16 (3) & \multicolumn{1}{l|}{A-1} & HABPanel-25 & 21 & $\times$ & $\times$ & $\times$ & 672 & 25 & 28 & 29 & 44 (3) & 19 (3) \\
\multicolumn{1}{l|}{R-6} & AnyMemo-440 & 90 & $\times$ & $\times$ & $\times$ & 950 & $\times$ & $\times$ & $\times$ & 342 (2) & 91 (3) & \multicolumn{1}{l|}{A-2} & Noad Player-1 & 8 & 10 & 5 & 12 & 5 & 6 & 10 & 10 & 13 (3) & 14 (3) \\
\multicolumn{1}{l|}{R-7} & Notepad-23 & 64 & 2719 & $\times$ & 1223 & 855 & $\times$ & 120 & 70 & $\times$ (0) & 116 (3) & \multicolumn{1}{l|}{A-3} & Weather-61 & 31 & $\times$ & $\times$ & $\times$ & $\times$ & 578 & 40 & 34 & 47 (2) & 16 (3) \\
\multicolumn{1}{l|}{R-8} & Olam-2 & 10 & $\times$ & 1288 & $\times$ & 101 & $\times$ & 10 & 10 & 62 (3) & 23 (3) & \multicolumn{1}{l|}{A-4} & Berkeley-82 & 8 & 50 & 683 & 39 & 22 & $\times$ & 8 & 8 & 41 (3) & 23 (3) \\
\multicolumn{1}{l|}{R-9} & Olam-1 & 10 & $\times$ & $\times$ & $\times$ & 37 & $\times$ & 10 & 10 & 24 (3) & 13 (3) & \multicolumn{1}{l|}{A-5} & andOTP-500 & 120 & $\times$ & $\times$ & $\times$ & $\times$ & $\times$ & 425 & 323 & 104 (3) & 103 (3) \\
\multicolumn{1}{l|}{R-10} & FastAdapter-394 & 8 & 526 & 10 & 1642 & 13 & 1506 & 8 & 8 & 61 (3) & 36 (3) & \multicolumn{1}{l|}{A-6} & K-9Mail-3255 & 14 & $\times$ & $\times$ & $\times$ & $\times$ & $\times$ & 15 & 19 & 45 (2) & 41 (3) \\
\multicolumn{1}{l|}{R-11} & LibreNews-22 & 93 & 1261 & $\times$ & $\times$ & $\times$ & $\times$ & 31 & 139 & 107 (1) & 42 (3) & \multicolumn{1}{l|}{A-7} & K-9Mail-3971 & 66 & $\times$ & $\times$ & $\times$ & $\times$ & $\times$ & 95 & 85 & $\times$ (0) & 152 (3) \\
\multicolumn{1}{l|}{R-12} & LibreNews-23 & 71 & $\times$ & 126 & 104 & 49 & 518 & $\times$ & 51 & $\times$ (0) & 83 (3) & \multicolumn{1}{l|}{A-8} & Firefox-3932 & 61 & $\times$ & $\times$ & $\times$ & $\times$ & $\times$ & 107 & 83 & $\times$ (0) & 47 (3) \\
\multicolumn{1}{l|}{R-13} & LibreNews-27 & 93 & 1272 & $\times$ & 197 & $\times$ & $\times$ & 31 & 139 & 11 (1) & 64 (3) & \multicolumn{1}{l|}{A-9} & Aegis-3932 & 117 & $\times$ & $\times$ & $\times$ & $\times$ & $\times$ & 159 & 129 & 205 (3) & 146 (2) \\
\multicolumn{1}{l|}{R-14} & SMSsync-464 & 118 & $\times$ & $\times$ & 443 & $\times$ & 63 & 186 & 254 & 213 (2) & 68 (3) & \multicolumn{1}{l|}{C-1} & NewPipe-7825 & 32 & $\times$ & $\times$ & $\times$ & $\times$ & $\times$ & 46 & 32 & 426 (1) & 64 (3) \\
\multicolumn{1}{l|}{R-15} & Transistor-63 & 41 & 94 & $\times$ & 216 & $\times$ & 17 & 10 & 48 & 151 (3) & 63 (3) & \multicolumn{1}{l|}{C-2} & SDBViewer-10 & 15 & $\times$ & $\times$ & 68 & 1900 & 36 & 20 & 192 & 98 (3) & 25 (3) \\
\multicolumn{1}{l|}{R-16} & Zom-271 & 50 & 536 & 45 & $\times$ & 50 & 24 & 52 & 59 & 120 (1) & 80 (3) & \multicolumn{1}{l|}{C-3} & Anki-10584 & 180 & $\times$ & $\times$ & $\times$ & $\times$ & $\times$ & 878 & 312 & 175 (2) & 148 (2) \\
\multicolumn{1}{l|}{R-17} & Pix-Art-125 & 48 & 1436 & $\times$ & 977 & 541 & 219 & 48 & 176 & 246 (3) & 34 (2) & \multicolumn{1}{l|}{C-4} & Alarmio-47 & $\times$ & $\times$ & $\times$ & $\times$ & $\times$ & 487 & $\times$ & $\times$ & A- (0) & A- (0) \\
\multicolumn{1}{l|}{R-18} & Pix-Art-127 & 56 & 1270 & $\times$ & 1416 & 1972 & 53 & 57 & 164 & 300 (1) & 49 (3) & \multicolumn{1}{l|}{C-5} & Shuttle-456 & 87 & $\times$ & $\times$ & $\times$ & $\times$ & $\times$ & 96 & 97 & 476 (1) & 132 (3) \\
\multicolumn{1}{l|}{R-19} & ScreenCam-25 & 33 & $\times$ & 152 & 28 & 43 & $\times$ & 40 & 33 & 69 (2) & 68 (3) & \multicolumn{1}{l|}{C-6} & Anki-3370 & 35 & $\times$ & $\times$ & $\times$ & $\times$ & $\times$ & 611 & 35 & 183 (2) & 41 (3) \\
\multicolumn{1}{l|}{R-20} & ownCloud-487 & 60 & 996 & 303 & 262 & $\times$ & 118 & 62 & 60 & 228 (1) & 87 (3) & \multicolumn{1}{l|}{C-7} & WhereUGo-368 & 165 & $\times$ & 2524 & $\times$ & $\times$ & $\times$ & 372 & 181 & $\times$ (0) & 157 (3) \\
\multicolumn{1}{l|}{R-21} & OBDReader-22 & 52 & $\times$ & 2525 & 1211 & $\times$ & $\times$ & $\times$ & 53 & 210 (2) & 76 (3) & \multicolumn{1}{l|}{C-8} & GrowTracker-87 & 210 & $\times$ & $\times$ & $\times$ & $\times$ & $\times$ & 271 & 605 & $\times$ (0) & 221 (3) \\
\multicolumn{1}{l|}{R-22} & Dagger-46 & 8 & 18 & 8 & 12 & 5 & $\times$ & 8 & 8 & 180 (1) & 14 (3) & \multicolumn{1}{l|}{C-9} & FakeStandby-30 & 27 & $\times$ & $\times$ & $\times$ & $\times$ & $\times$ & 27 & 27 & 386 (1) & 38 (3) \\
\multicolumn{1}{l|}{R-23} & ODK-2086 & 640 & 2064 & $\times$ & 1538 & 531 & 725 & 939 & 1016 & 180 (1) & 41 (3) & \multicolumn{1}{l|}{C-10} & getodk-219 & 20 & 45 & 166 & 29 & 32 & $\times$ & 157 & 20 & 80 (3) & 25 (3) \\
\multicolumn{1}{l|}{R-24} & K-9Mail-3255 & 13 & $\times$ & $\times$ & $\times$ & $\times$ & $\times$ & 16 & 18 & 256 (3) & 43 (3) & \multicolumn{1}{l|}{} &  &  &  &  &  &  &  &  &  &  & \multicolumn{1}{l}{}  \\ \bottomrule

\end{tabular}
}
\label{tab:expe_all}
\vspace{-0.15in}
\end{table*}

Table \ref{tab:expr_time_stat} and Table \ref{tab:expe_all} show the reproducing time of {\tool} and four baselines on the successfully reproduced reports. 
For the 46 crash reports that {\tool} can reproduce, the average reproducing time of {\tool} is 68.7 seconds, which indicates that {\tool} can automatically reproduce crashes based on the corresponding stack trace within an acceptable time cost. 
Compared with the baselines, {\tool} performs better than 
ReCDroid$_{ST}$ and
the automated GUI testing techniques in most of the cases. 
Since different tools can succeed in different crashes, for the average reproducing time, we compare {\tool} with each baseline on the crashes that both of them can reproduce. 
The results show that {\tool} is 
1110\% faster than ReCDroid$_{ST}$ (81.6s vs. 988s),
1611\% faster than Monkey (39.1s vs. 669s), 620\% faster than Humaniod (73.9s vs. 532s), 705\% faster than Ape (66s vs. 531s), and 302\% faster than Q-testing (89.6s vs. 360s).

\subsection{RQ2: Contribution of Different Scorers}
Columns CT$_{wp}$ and CT$_{ww}$ in Table \ref{tab:expe_all} show the reproduction results for the variants of {\tool} without the page reaching scorer and without the widget hitting scorer, respectively. Overall, CT$_{wp}$ can only successfully reproduce 42 crash reports, 4 fewer than {\tool}. Meanwhile, the average reproducing time of CT$_{wp}$ is 127.1 seconds, which is 82\% slower than {\tool} (69.7s vs. 127.1s) considering the reports both can reproduce. 
For CT$_{ww}$, it can successfully reproduce 44 crashes (2 fewer than {\tool}), and the average reproducing time is 113.2s (63\% slower than {\tool}). 
The inferior performance of CT$_{wp}$/CT$_{ww}$ indicates that the two scorers can significantly improve the effectiveness and efficiency of crash reproducing. 

We further examine the detailed difference between the results of CT$_{wp}$/CT$_{ww}$ and {\tool} for a thorough understanding. 
For the reports which involve shorter exploration sequences for reaching the crash-occurring page, e.g., \textit{R-10 FastAdapter-394}, we find that excluding the page reaching scorer or widget hitting scorer would not largely influence the reproduction.
This might be because, in these cases, the widget transferring to the crash-occurring page can be found by one of the scorers or even by traversal.
By comparison, for the reports which involve longer exploration sequences for reproduction (e.g., \textit{A-5 andOTP-500}), or the entry page has dozens of candidate widgets (e.g., \textit{R-5 AnyMemo-18}), it is difficult to find the reproducing steps solely by the traversal or the widget hitting scorer. 
In this case, the page reaching scorer which is accomplished by LLM contributes significantly to the crash reproduction by providing step-by-step guidance to the crash-occurring page. 
Besides, for the reports which have many candidate widgets on the crash-occurring page (e.g., \textit{R-14 SMSsync-464}), the efficiency can be largely improved by the widget hitting scorer that predicts which widgets should be interacted with to trigger the crash.

Still, we need to admit that on some reports like \textit{R-11 LibreNews-22}, removing the scorer would improve the crash reproduction efficiency. 
This is mainly because the scorers can occasionally fail to conduct an accurate prediction, and with the wrong guidance, the exploration might go in the wrong direction and waste time.

\subsection{RQ3: Usefulness of {\tool}}
Columns P$_{trace}$ and P$_{step}$ of Table \ref{tab:expe_all} show the results of participants' manual reproduction based on stack traces and reproducing steps (generated by {\tool}), respectively. 
Following the reproducing steps generated by {\tool}, 100\% of reports are successfully reproduced by at least two participants.
As a comparison, when we only provide the stack trace, this percentage drops to 63\% (29/46), and there are 6 reports that none of the participants can reproduce.
Besides, the average reproducing time with {\tool} is 66.7 seconds, which is 215\% faster than reproducing from stack traces (57.3s vs. 180.5s) considering the reports that both have at least one participant can reproduce.
This indicates the usefulness of our proposed {\tool}, whose generated reproducing steps can supplement the stack-trace-only crash reports and make the crashes easily to be reproduced.

Furthermore, on 91.3\% (42/46) reports, the reproduction is faster for the automatic approach {\tool}, when compared with the average time of human reproduction directly from the stack trace. This further implies the usefulness of {\tool} which provides an automatic solution and can be faster than humans.


After practitioners have manually reproduced the bug reports, we also conduct an unstructured interview about the challenges they encountered in reproducing crashes based on stack traces only.  
Most participants complain that crash-related information in stack traces is too limited to infer how to reproduce crashes. 
They usually need to spend a long time trying to reach crash-occurring pages and finding crash-triggering widgets, and after many attempts, they may lose interest and assume that crashes are not reproducible. 
By comparison, they agree that {\tool} can automatically conduct the tedious process of exploring the app and finding paths to crash-occurring pages and crash-triggering widgets.


\section{DISCUSSION}
\label{sec_discussion}

\subsection{Limitations}
\label{subsec_dislim}

Except for the two engineering limitations discussed in Section \ref{sec_result_rq1}, there are three other technical limitations of {\tool} which hinder it from reproducing all bug reports, 
This also indicates the challenges in crash reproduction from stack traces and calls for further research.

First, {\tool} may fail to reproduce crashes that do not contain crash-occurring pages or crash-involved APIs in stack traces. In some special cases, crashes do not occur in a specific activity or fragment, e.g., faults related to network communication.
For these cases, crash-occurring pages and crash-involved APIs are not available from the stack trace, {\tool} may degenerate into an automated GUI testing tool for aimless exploration due to the lack of guidance from the stack trace.

Second, {\tool} cannot reproduce crashes requiring valid input contents, e.g., performing a crash-triggering log-in requires a valid username and password. Such crashes require human intervention and cannot be fully automated since input contents are not present in the stack trace.
Nevertheless, by asking the users to provide the information in advance, {\tool} can still conduct the automatic reproduction for these cases. 

Third, {\tool} cannot capture special preconditions that trigger crashes. For example, a display GUI page works fine with default settings, but if switching to a dark theme, a crash would occur when reaching the page. In this case, {\tool} can correctly understand that the display page is a crash-occurring page, but it could not successfully reproduce the crash even if the correct page is reached. 
This is because {\tool} can hardly capture the triggering precondition of \textit{switching to the dark theme} from the stack trace.
In the future, we will investigate incorporating the crash-triggering preconditions into {\tool} with the help of static code analysis and other techniques.


\subsection{Threats to Validity}


The first threat relates to the randomness of {\tool}.
Our widget selector (Section \ref{subsubsec_selector}) will select a non-optimal widget with a certain probability. To reduce this threat, we run {\tool} and its variants (CT$_{wp}$ and CT$_{ww}$) three times and record the average reproducing time in experiment results. 

The second threat relates to the choice of parameter settings of {\tool} that may affect the effectiveness and efficiency of crash reproduction. 
In order to mitigate the threat, we conduct small-scale experiments on several ``crash'' reports that are excluded from our experimental data to determine suitable settings before the evaluation. 
These reports come from the data collection (Section \ref{sec_movitation}) when we find some ``crash'' reports with stack traces, but they can not be triggered by 
ourselves due to incompatibility with our environment.


The third threat relates to the confounding effects of participants. Following the existing approach \cite{zhao2019ReCDroid}, we assume that students with Android programming experience can be substituted for testers, and their reproducing time and success rate are representative.





\section{RELATED WORK}
\label{sec_related_work}

\textbf{Mobile Bug Reports Analysis and Reproducing. }There are several studies which utilize natural language processing techniques to extract critical information from mobile bug reports, such as summarizing and classifying bug reports \cite{gegick2010identifying, rastkar2014automatic}, facilitating dynamic analysis \cite{jin2014preffinder, wong2015dase}, augmenting bug reports for mobile apps \cite{moran2015auto, Liu2020Automated} and generating test cases \cite{moran2015auto, chaparro2019assessing}.

Several studies focus on crash reproduction of mobile bug reports with step-by-step guidance, i.e., textual described reproducing steps \cite{zhao2019ReCDroid, zhao2022recdroid+, zhang2023automatically} and visual recordings \cite{feng2022gifdroid}. Specifically, ReCDroid \cite{zhao2019ReCDroid} and ReCDroid+ \cite{zhao2022recdroid+} leveraged the natural language described reproducing steps to perform reproduction. It designed a set of predefined grammar patterns to extract events and objects from textual reproducing steps and then adopted a greedy-based dynamic exploration to synthesize event sequences. 
ReproBot \cite{zhang2023automatically} went a step further by analyzing the reproducing steps more accurately and designing a new exploration strategy to find the best match between steps and GUI actions. 
GIFdroid \cite{feng2022gifdroid} leveraged the visual screen record to perform reproduction. 
It adopted image-processing techniques to map the keyframes in recording to GUI states and generated reproducing traces based on the transition graph. 
Compared to the aforementioned crash reproduction tools, CrashTranslator achieves the following advances: 1) Our work enables the reproduction of crashes from stack traces without relying on step-by-step guidance. This is a more challenging task, and existing tools perform poorly; 2) We propose a novel approach which leverages LLM and reinforcement learning to predict and guide the exploration steps for trigger the crash, which is more effective and efficient than previous techniques. 

There are also studies which record and replay bugs in mobile and web apps by using running information \cite{nurmuradov2017caret, yan2021efficient, gomez2013reran, ko2006barista, white2015generating} and textual description \cite{bell2013chronicler, kifetew2014reproducing}. 
Among these studies, CrashDroid \cite{white2015generating} generated reproducing steps by translating the call stack, which contains all method calls from app launch to app crash.
It requires the call stack collected by the specific mechanism throughout the app's run, while our approaches only need the automatically generated stack trace when the crash occurs. 


\textbf{Stack Trace Analysis. }Stack traces offer exception-related information about an app. Schroter et al. \cite{schroter2010stack} empirically indicated that the stack trace information was very helpful to developers when debugging. Subsequently, several automatic approaches were proposed to recover the links between the crashes and their cause functions and assist developers in locating crashing faults \cite{wu2014crashlocator, wong2014boosting, moreno2014use, gong2014locating, gu2019does}. Going a step further, researchers started to explore how to utilize the located faulty functions to help developers fix bugs, e.g., generating test cases for fault functions \cite{rossler2013reconstructing, chen2014star, xuan2015crash, soltani2016evolutionary, soltani2017guided, soltani2020benchmark}, and finding the best developer to fix the bug \cite{kim2011crash, sushentsev2022dapstep}. Besides, there were some studies focused on calculating the similarity between the stack trace, which could be used to distinguish duplicate crash reports \cite{rodrigues2022fast, dang2012rebucket, rodrigues2022tracesim, khvorov2021s3m, vasiliev2020tracesim}.
This study opens a new direction, i.e., the crash reproduction directly from the stack trace.


\section{Conclusion}


Crash reports from open-source platforms are vital for ensuring mobile application quality. Still, the crash-related information they record is not always complete, e.g., they may only contain the crash stack trace but lack the reproducing steps, which hinders developers from fixing issues. This paper proposes a novel reproduction approach named {\tool} which automatically reproduces crashes from stack traces of mobile bug reports. It adopts three scorers, i.e., page reaching scorer, widget hitting scorer, and exploration optimization scorer, to select crash-related widgets iteratively until the target crash is triggered. We evaluate the {\tool} on 75 bug reports, and it can successfully reproduce 46 (61.3\%) of the crashes within an acceptable time cost, largely outperforming automated GUI testing baselines.

\section{ACKNOWLEDGMENTS}
This work was supported by the National Natural Science Foundation of China Grant No.62232016, No.62072442 and No.62272445, and Youth Innovation Promotion Association Chinese Academy of Sciences.

\bibliographystyle{ACM-Reference-Format}
\bibliography{reference}

\end{document}